\documentclass[pra,twocolumn,showpacs]{revtex4}

\usepackage{graphicx}
\usepackage[bookmarks=true]{hyperref}
\usepackage{amsmath}
\usepackage{amssymb}

\newcommand{\Idf}{\leavevmode\hbox{\scriptsize1\footnotesize\kern-.36em1}}
\newcommand{\carb}{$^{13}$C }
\newcommand{\ket}[1]{\vert{#1}\rangle}
\newcommand{\bra}[1]{\langle{#1}\vert}
\newcommand{\tr}[1]{\textrm{Tr}\left\{{#1}\right\}}
\newcommand{\braket}[2]{\langle{#1}\vert{#2}\rangle}
\newcommand{\half}{\frac{1}{2}}
\newcommand\Id{\leavevmode\hbox{\small1\normalsize\kern-.33em1}}
\newcommand{\ham}{{\mathcal{H}}}
\newcommand{\kpsi}{\vert{\psi}\rangle}

\graphicspath{{./Figures/},{../../../figures/FAP/},{../../../Mathematica/FAP/},{../../../figures/FAP/NewTransport/}}
\renewcommand{\emph}{\textit}
\begin{document}

\title{Coherent state transfer via highly mixed quantum spin chains}

\author{Paola Cappellaro}
\affiliation{Nuclear Science and Engineering Department, Massachusetts
Institute of Technology, Cambridge, Massachusetts 02139, USA}
\author{Lorenza Viola} 
\affiliation{\mbox{Department of Physics and Astronomy, Dartmouth College, 
6127 Wilder Laboratory, Hanover, New Hampshire 03755, USA}}
\author{Chandrasekhar Ramanathan}
\affiliation{\mbox{Department of Physics and Astronomy, Dartmouth College, 
6127 Wilder Laboratory, Hanover, New Hampshire 03755, USA}}

\begin{abstract}
Spin chains have been proposed as quantum wires in many quantum
information processing architectures. Coherent transmission of quantum
information over short distances is enabled by their internal
dynamics, which drives the transport of single-spin excitations in
perfectly polarized chains.  Given the practical challenge of
preparing the chain in a pure state, we propose to use a chain that is
initially in the maximally mixed state. We compare the transport
properties of pure and mixed-state chains, finding similarities that
enable the experimental study of pure-state transfer by its simulation
via mixed-state chains, and demonstrate protocols for the perfect
transfer of quantum information in these chains. Remarkably,
mixed-state chains allow the use of Hamiltonians which do not preserve
the total number of excitations, and which are more readily obtainable
from the naturally occurring magnetic dipolar interaction.  We propose
experimental implementations using solid-state nuclear magnetic
resonance and defect centers in diamond.
\end{abstract}

\maketitle

%====================================================
\section{Introduction}
%====================================================

Many quantum information processing (QIP) proposals require the
computational units to be spatially separated due to constraints in
fabrication or control~\cite{Burgarth06, Jiang08b,Campbell08}.
Coherent information transfer from one quantum register to another
must then be carried out either by photons or, for more compact
architectures, by quantum wires. Linear chains of spins have been
proposed as quantum wires, the desired transport being obtained via
the free evolution of the spins under their mutual
interaction~\cite{Bose03,Christandl04,Christandl05,Cappellaro07l}. In
general, only partial control over the spins in the chain is assumed,
as relevant to most experimental implementations, and perfect state
transfer with no or reduced control requirements has already been
studied~\cite{Fitzsimons06,Burgarth10}. The reduced control may also
naturally entail an imperfect initialization of the spin chains, a
constraint addressed in some recent
work~\cite{DiFranco08c,Markiewicz09}. While much of the literature on
spin chains has focused on transport in the first excitation manifold,
imperfect chain initialization makes it imperative to study the
transport properties of the higher excitation manifolds and, more
generally, of mixed-state spin chains.
 
In this paper we focus on the transport properties of chains that are
initially in the \textit{maximally mixed-state}.  This state
correspond to the infinite temperature limit and is easily reachable
for many systems of relevance to QIP~\cite{Wang04, Lao04,Ruess07}.
Alternatively, it could be obtained by an active randomization of the
chain's initial state.
The reduced requirements on the initialization of the wires, when
combined with low control requirements, would make quantum information
transport more accessible to experimental implementations.  We are
thus interested in comparing the transport properties of pure and
mixed state chains, with a twofold goal in mind: i) exploring the
extent to which the experimental study of pure-state transport may be
enabled by its simulation via highly mixed chains; and ii) studying
protocols for the transport of quantum information via mixed-state
chains.

The paper is organized as follows. We first review in Sec. II some
results about transport in the first excitation manifold and then
generalize them to higher excitation manifolds and mixed
states. Furthermore, we describe how transport may also be driven by
Hamiltonians that do not conserve the excitation number.  In Sec. III,
we investigate transfer of quantum information in a mixed-state chain
based on a standard encoding protocol and extend it to more general
Hamiltonians. In Sec. IV we then present applications of these
results, focusing on two experimental QIP platforms. The first
is based on solid-state nuclear magnetic resonance (NMR) and enables
the study of transport in mixed state chains, and its limitations due
to imperfections in the system. The second example is an application
of quantum information transfer via mixed-state wires in a scalable
architecture based on spin defects in diamond.

%=================================================================
\section{State transfer in pure- and mixed-state spin chains}
%=================================================================
\subsection{Single-spin excitation manifold}
%====================================================

In analogy with the phenomenon of spin waves, the simplest mechanism
for quantum state transfer 
is the propagation of a single spin excitation
$\ket{j}=\ket{00\dots01_j0\dots}$ down a chain of $n$ spins-1/2,
coupled by the Heisenberg exchange Hamiltonian~\cite{Bose03,Bayat10}.
In this context, the most common model studied is the xx-model,
described by the Hamiltonian
\begin{equation}
		\ham_{\textsf{xx}}=\sum_{j=1}^{n-1}\frac{d_{j}}2(\sigma_x^j\sigma_x^{j+1}
		+ \sigma_y^j\sigma_y^{j+1}),
\label{eq:Hxx}
\end{equation}
where $\sigma_\alpha$ ($\alpha=\{x,y,z\}$) are the Pauli matrices and
we have set $\hbar=1$. A single spin excitation is propagated through
the chain via energy conserving spin flip-flops, as shown by rewriting
the xx-Hamiltonian in terms of the operators $\sigma_\pm=(\sigma_x\pm
i\sigma_y)/2$:
\begin{equation}
		\ham_{\textsf{xx}}=\sum_{j=1}^{n-1}d_{j}(\sigma_+^j\sigma_-^{j+1}
		+ \sigma_-^j\sigma_+^{j+1}).
\label{eq:Hxy1}
\end{equation}
The transport property of the xx-Hamiltonian are made apparent by a
mapping of the system to a local fermionic Hamiltonian via the
Jordan-Wigner transformation:
\begin{equation}\begin{array}{cc}
 c_j=\displaystyle\prod_{k=1}^{j-1}(-\sigma^k_z)~ \sigma^j_-,\ \ &
\sigma^j_-=\displaystyle\prod_{k=1}^{j-1}\left(1/2- c_k^\dag c_k
\right)c_j,
\end{array}
\label{eq:fermion}
\end{equation}
which also yields $\sigma_z^j=1-2c_j^\dag c_j$. Using these fermion
operators, the xx-Hamiltonian can be rewritten as:
\begin{equation}
	\ham_{\textsf{xx}}=\sum_{j=1}^{n-1}d_j (c_j^\dag c_{j+1}
	+c_{j+1}^\dag c_j ).
\label{eq:HxyFerm}
\end{equation}
Since the total angular momentum along $z$,
$Z=\sum_{j=1}^n\sigma_z^j$, is conserved, $[\ham_{\textsf{xx}},Z]=0$,
it is possible to block-diagonalize the Hamiltonian into subspaces
corresponding to (typically degenerate) eigenvalues of ${Z}$. These
subspaces are more simply characterized by the number of spins in the
excited state $\ket{1}$, which is usually called the (magnon)
excitation number.  In this description, the xx-Hamiltonian induces
transport by creating an excitation at site $j+1$ while annihilating
another at site $j$.  For a given evolution time $t >0$, transport
from spin $j$ to spin $l$ is characterized by the transfer fidelity of
the state $\ket{j}$ to $\ket{l}$, defined as the overlap
$P_{jl}^{\textsf{xx}}(t)=|A_{jl}(t)|^2=|\bra{l}U_{\textsf{xx}}(t)\ket{j}|^2$,
where $U_{\textsf{xx}}(t)=e^{-i\ham_{\textsf{xx}}t}$ and usually $j=1$
and $l=n$ in a open-ended chain.

A well studied case~\cite{Christandl04,Christandl05,Cappellaro07l} is
the \textit{homogenous} limit, corresponding to equal couplings,
$d_j=d$ for all $j$.  The corresponding Hamiltonian can be
diagonalized by the operators
\begin{equation}
\label{eq:EqualAop}
a_k^h=\frac{1}{\sqrt{n+1}}\sum_{j=1}^n\sin{(\kappa j)}\,c_j,\ \ \ 
\kappa=\frac{\pi k}{n+1}, \;k=1,\ldots,n,
\end{equation}
to reveal the eigenvalues $\omega^h_k=2\,d\cos{(\kappa)}$.  It is then
possible to calculate the probability of state transfer from spin $j$
to spin $l$, yielding ${P}^{h,\textsf{xx}}_{jl}(t)=|
A^{h}_{jl}(t)|^2$, with~\cite{Christandl04}:
\begin{equation}	
{A}^{h}_{jl}(t)=
\frac{2}{n+1}\sum_k\sin{(\kappa j)}\sin{(\kappa l)}e^{-i\omega_kt}.
	\label{eq:poltransfer}
\end{equation}

In practice, it is often difficult to experimentally prepare the spins
in the maximally polarized, ground state. Thus, in order to
experimentally investigate quantum transport it is highly desirable to
relax the requirements on the initial state of the spin chain. In
\cite{Cappellaro07l}, we found that it was possible to simulate the
spin excitation transport by using a highly mixed spin chain.  We
generalized the spin excitation transport to mixed states by looking
at the evolution of an initial state of the form
$$\rho=\frac{1}{2^n}(\openone+\epsilon\,\delta\rho_z^j), \;\;\;
\delta\rho_z^j=\openone_{j-1}\otimes\sigma_z^j\otimes\openone_{n-j}.$$
\noindent 
This state represents a completely mixed-state chain with a single
spin partially polarized along the $z$-axis.  Notice that we only need
to follow the evolution of the traceless deviation $\delta\rho$ from
the identity, since it is the only non-trivial part as long as the
dynamics is unital.  The goal is then to transfer the state
$\delta\rho_z$ from spin $j$ to spin $l$.  A metric describing the
transfer efficiency is the correlation of the evolved state with the
target final state $\delta\rho_z^l$,
$C_{jl}(t)=\tr{\delta\rho_z^j(t)\delta\rho_z^l}$.  Using a fermionic
mapping of the mixed states, we found in Ref.~\cite{Cappellaro07l}
that for the homogenous xx-Hamiltonian such a correlation is
\textit{exactly} given by ${P}^{h,\textsf{xx}}_{jl}(t)$, although the
states involved in the transport are quite different. Indeed, states
such as $\delta\rho_z^j$ do not reside in the lowest excitation
manifold, for which the state transfer equation (\ref{eq:poltransfer})
was initially calculated, but they are a mixture spanning all the
possible excitation manifolds.

A similar mapping from mixed to pure states cannot be carried further
in such a simple way. For example, we cannot use the state
$\delta\rho_x^j=\openone_{j-1}\otimes\sigma_x^j\otimes\openone_{n-j}$
to simulate the transfer of a coherent pure state such as
$\ket{+}\ket{00\dots}$, where $\ket{+}=(\ket{0}+\ket{1})/\sqrt{2}$.
In the following, we will analyze the conditions allowing state
transfer in mixed-state spin chains in order to lay the basis of a
protocol for the transport of quantum information.

%======================================================================
\subsection{Evolution in higher excitation manifolds}
%======================================================================

Since highly mixed states include states with support in all the spin
excitation manifolds, we first analyze the evolution of higher
excitation energy eigenstates.  Thanks to the fact that it conserves
the excitation number, the xx-Hamiltonian [Eq. (\ref{eq:Hxx})] is
diagonal in each excitation subspace.  Let the eigenstates in the
first excitation subspace be denoted by $\ket{E_k}$ (e.g.,
$\ket{E_k}=\sqrt{\frac{2}{n+1}}\sum_j \sin{(\kappa j)}\ket{j}$ in the
homogeneous case).  Since the xx-Hamiltonian describes non-interacting
fermions, eigenfunctions of the higher manifolds can be exactly
expressed in terms of Slater determinants of the one-excitation
manifold.  Consider for example the case of the 2-excitation manifold,
described by states $\ket{pq}=\ket{0...1_p..0..1_q...0}$. The
eigenstates $\ket{E_{kh}}$ are
\begin{equation}
\ket{E_{kh}}=
\half\sum_{pq}\big(\braket{E_k}{p}\braket{E_h}q-\braket{E_k}q
\braket{E_h}p\big)\,\ket{pq},
\end{equation}
with eigenvalues $E_{kh}=E_k+E_h$.  We can then calculate the
time evolution as
\begin{equation}
\begin{array}{c}
U_{\textsf{xx}}(t)\ket{pq}=\sum_{k,h}e^{-i(\omega_k+\omega_h)t}
\braket{E_{kh}}{pq}\braket{rs}{E_{kh}}
\ket{rs}\\\qquad\qquad=\sum_{r,s} A_{pq,rs}(t) \ket{rs}, 
\label{eq:}
\end{array}
\end{equation}
where
\begin{equation} 
A_{pq,rs}(t)=\left| \begin{array}{cc} A_{pr}(t)&A_{ps}(t)\\
A_{qr}(t)&A_{qs}(t)\end{array}\right|,
\end{equation}
and $A_{pr}(t)$ describes the amplitude of the transfer 
in the one-excitation manifold,
$A_{pr}(t)=\bra{r}U_{\textsf{xx}}(t)\ket{p}$.

More generally, for an arbitrary initial eigenstate of the total
$z$-angular momentum, $\ket{\vec{p}}=\ket{p_1,p_2,\dots}$, with
$p_k\in\{0,1\}$, the transfer amplitude to the eigenstate
$\ket{\vec{r}}$ is given by
\begin{equation} 
A_{\vec{p}\,\vec{r}}(t)=\left| \begin{array}{ccc}
A_{p_1r_1}(t)&A_{p_1r_2}(t)&\dots\\
A_{p_2r_1}(t)&A_{p_2r_2}(t)&\dots\\\dots&\dots&\dots\end{array}\right|.
\end{equation}
We can then calculate the transfer of any initial mixed state
$\rho_a=\sum_{\vec{p},\vec{q}}
a_{\vec{p}\,\vec{q}}\ket{\vec{p}}\bra{\vec{q}}$ to another mixed state
$\rho_b$, as $M_{ab}(t)=\sum b_{\vec{r}\vec{s}}\, a_{\vec{p}\,\vec{q}}
A_{\vec{p}\,\vec{r}}(t) A_{\vec{q}\,\vec{s}}^*(t)$~\endnote{For some
states having a simple form in fermionic operators, it might instead 
be advantageous to calculate the transport correlation functions
directly~\cite{Cappellaro07l,Feldman96}.}.

It is important to stress that the above expressions allow us to
calculate the evolution of any mixed state for any choice of couplings
in the xx-Hamiltonian of Eq. (\ref{eq:Hxx}), as we only used the
property that this Hamiltonian describes non-interacting fermions.
Thus, the higher excitations are seen to propagate simultaneously at
the same group velocity. This result can be used to search for
coupling distributions that give better state transfer properties than
the equal-coupling case. In particular, because the transfer of the
one-spin polarization state $\delta\rho_z^j$ is found to have the same
expression as the spin-excitation state transfer, we can use known
results for the latter to find optimal coupling distributions.

%============================================================================
\subsection{Perfect state transfer for engineered Hamiltonians}\label{Perfect}
%=============================================================================

\begin{figure*}[hbt]
	\centering
		\includegraphics[scale=0.64]{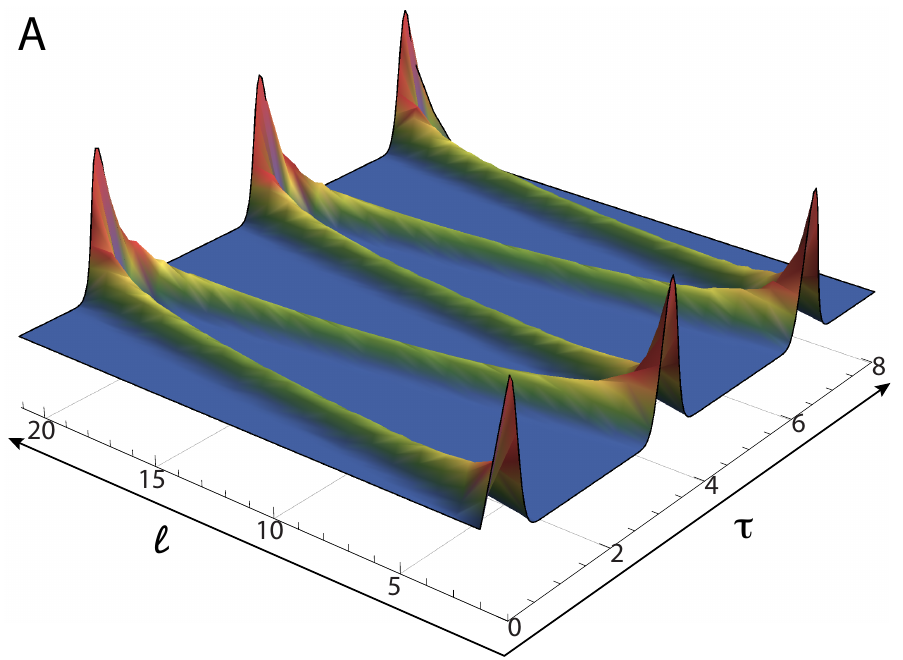}\ \ \ \ \ \ \ \ 
		\includegraphics[scale=0.64]{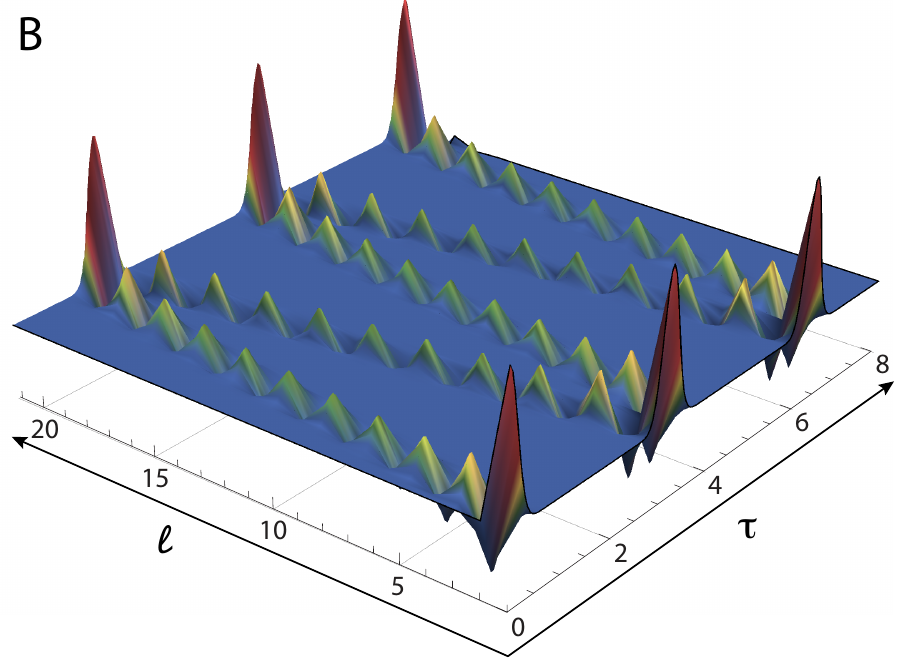}
\caption{(Color online) (A) Transport of polarization under the
xx-Hamiltonian with optimal couplings, Eq. (\ref{optcoup}). Shown is
the intensity of the polarization at each spin site
$P_{1,\ell}^{o,\textsf{xx}}(t)$ as a function of normalized time
$\tau=4dt/n$ for a propagation starting from spin 1. The chain length
$n=21$ spins. (B) Transport of polarization under the dq-Hamiltonian
$P_{1,\ell}^{o,\textsf{dq}}(t)$, Eq. (\ref{eq:poltransferdq}), with
the same parameters as in (A).}
	\label{TrPolFF}
\end{figure*}

Although spin excitations propagate through the chain for any
xx-Hamiltonian, as seen in the homogeneous case this does not always
allow for perfect state transfer because of wave-packet
dispersion~\cite{Osborne04,Ramanathan10}.  Good transport properties
have been found for a class of Hamiltonians that have been suitably
engineered, either by modifying the coupling strengths among the spins
or by introducing an additional spatially varying magnetic
field~\cite{Christandl05,Difranco08a}.
In particular, the Hamiltonian
\begin{equation}
\ham^o_{\textsf{xx}}=\sum_{j=1}^{n-1}2d\sqrt{\frac{j(n-j)}{n^2}}(\sigma_+^j\sigma_-^{j+1}
	+ \sigma_-^j\sigma_+^{j+1})
\label{optcoup}
\end{equation}
allows for optimal transport of the excitation from the first to the
last spin in the chain.  Not only does this choice of couplings allow
for perfect
transport~\cite{Nikolopoulos04,Nikolopoulos04b,Christandl05} but it
does so in the shortest time~\cite{Kay06,Yung06}.  Notice that in
Eq. (\ref{optcoup}) we expressed the couplings in terms of the
maximum coupling constant $d$, since typically this will be
constrained in experimental implementations, as opposed to the more
common choice in the literature, whereby
$d_j=\frac{d'}2\sqrt{j(n-j)}$, with $d'=4d/n$.

The optimal coupling Hamiltonian $\ham^o_{\textsf{xx}}$ can be
diagonalized by the following fermion operators~\cite{Shore78,Cook79}:
\begin{equation}
\begin{array}{l}
	a^o_k = \sum_j\alpha_j(k)c_j,\\
	\alpha_j(k)=\frac{2^{\frac{n+1}{2}} }{2^j} \sqrt{\frac{k}{j}
	\binom nk/\binom nj} J_{n-j}^{(j-k,j+k-n-1)},
	\end{array}
\label{eq:fermo}
\end{equation}
where $J_n^{(a,b)}$ is the Jacobi polynomial evaluated at 0. The
eigenvalues read $\omega^o_k=\frac{2d}n[2k-(n+1)]$.  The transfer
amplitude $A_{jl}^{o,\textsf{xx}}(t)$ between spin $j$ and spin $l$
then becomes
$A_{jl}^{o,\textsf{xx}}(t)=\sum_{k}\alpha_j(k)\alpha_l(k)e^{-i\omega^o_kt}$,
which yields the transfer function $P_{jl}^{o,\textsf{xx}}(t)=
|A_{jl}^{o,\textsf{xx}}(t)|^2$.  Using these results, we can calculate
the transfer probability from spin 1 to spin $n$ of the one-spin
excitation in a pure-state chain,
\begin{equation}
P^{o,\textsf{xx}}_{1n}=\left[\sin(\tau)\right]^{2(n-1)}, \;\;\;\;
\tau=\frac{4d\,t}n.
\end{equation}  
The same expression also describes the transport of the
spin-polarization ($\delta\rho_z^j$) in a mixed-state chain. Notice
that at a time $t^\star=\frac{\pi}{2}\frac{n}{4d}$, perfect transfer
is achieved. This optimal time reflects the maximum speed of the
transport, which is given by the group velocity, $v_g=4d\frac{2}\pi$,
of the spin wave traveling through the
chain~\cite{Osborne04,Ramanathan10}.

Perfect state transfer is achieved not only for the choice of
couplings in Eq. (\ref{optcoup}) but, more generally, for a class of
xx-Hamiltonians that support either a linear or a quadratic
spectrum~\cite{Albanese04,Shi05,Kostak07,Xi08}. It was observed in
fact that these Hamiltonians allow for perfect mirror inversion of an
arbitrary (pure) input state. A different approach to perfect state
transfer, with a generic Hamiltonian spectrum, is to confine the
dynamics of the system to an effective two-qubit
subspace~\cite{Gualdi08,Gualdi09}, which by construction is always
mirror-symmetric. The confinement is obtained by weakening the
couplings of the first and last qubit in the chain. A similar approach
could be taken to achieve perfect transfer with mixed-state chains
using the Hamiltonian in Eq. (\ref{eq:Hxx}) (or the Hamiltonian that
will be discussed in the next section, Eq. (\ref{eq:Hdq})) provided
that $d_1,d_{n-1}~\ll~d_i$.  For more general long-range Hamiltonians,
such as the XXZ
dipolar Hamiltonian considered in~\cite{Gualdi08,Gualdi09}, the
equivalence of the evolution between pure and mixed state is lost and
it is thus not possible to directly apply this strategy.

%===========================================================
\subsection{Transport via double-quantum Hamiltonian}
%===========================================================

In the previous sections we showed that the transport features of
xx-Hamiltonians relied on the mapping to free fermions and the
conservation of excitation number. It is therefore surprising to find
another class of Hamiltonians that show very similar transport
properties even if they do not conserve the excitation
number. Consider the so-called double-quantum (dq) Hamiltonian
\begin{equation}
		\ham_{\textsf{dq}}=\hspace*{-.7mm}\sum_{j}\frac{d_{j}}{2}(\sigma_x^j\sigma_x^{j+1}
		-
		\sigma_y^j\sigma_y^{j+1})=\hspace*{-.7mm}\sum_{j}d_{j}(\sigma_+^j\sigma_+^{j+1}
		+ \sigma_-^j\sigma_-^{j+1}).
\label{eq:Hdq}
\end{equation}
As this Hamiltonian does not conserve the excitation number,
$[\ham_{\textsf{dq}},Z]\neq0$, we would not expect it to support the
transport of single-spin excitations. However, as observed
in~\cite{Cappellaro07l,Doronin00}, the dq-Hamiltonian is related to
the xx-Hamiltonian by a simple similarity transformation,
$U^{\textsf{xx}}_{\textsf{dq}}=\prod_j\sigma^{2j+1}_x$. Therefore, the
dq-Hamiltonian commutes with the operator $\tilde{Z}=\sum_j
(-1)^{j+1}\sigma_z^j$ and it can be block-diagonalized following the
subspace structure defined by the (degenerate) eigenvalues of
$\tilde{Z}$.  The dq-Hamiltonian allows for the mirror inversion of
states contained in each of the subspaces defined by the eigenvalues
of $\tilde{Z}$ (the equivalent of single-spin excitation and higher
excitation manifolds for $Z$).  For pure states, these states do not
have a simple interpretation as local spin excitations, and the
dq-Hamiltonian is thus of limited practical usefulness for state
transfer. Interestingly, however, the situation is more favorable for
the transport of spin polarization in mixed-state chains. Indeed,
states such as $\delta\rho_z^j$ are invariant, up to a sign change,
under the similarity transformation
$U^{\textsf{xx}}_{\textsf{dq}}$. Thus we can recover the results
obtained for the polarization transport under the xx-Hamiltonian for
any coupling distribution:
\begin{equation}	
{P}^{\textsf{dq}}_{jl}(t)=(-1)^{j-l}| A_{jl}^{\textsf{xx}}(t)|^2.
	\label{eq:poltransferdq}
\end{equation}
In figure \ref{TrPolFF} we illustrate the transport of polarization
from spin $j=1$ as a function of the spin number $\ell$ and time.
Comparing figure \ref{TrPolFF}(A) with figure \ref{TrPolFF}(B), that
show the transport under the optimal coupling xx- and dq-Hamiltonian
respectively, we see enhanced modulations due to the positive-negative
alternation of the transport on the even-odd spin sites. Despite this
feature, perfect transport is possible even with the dq-Hamiltonian,
which, unlike the xx-Hamiltonian, can be easily obtained from the
natural dipolar Hamiltonian with only collective
control~\cite{Munowitz87b,Ramanathan03}.

%==============================================================================
\section{Protocol for mixed state quantum information transport}\label{qubit}
%==============================================================================

In the previous section we showed that mixed-state chains have
transport properties similar to pure-state chains. However, while a
pure eigenstate of the $Z$ operator is transported using a mixed-state
chain, coherences are not.  This means that it is possible to transfer
a bit of \textit{classical} information by encoding it in the
$\ket{0}$ and $\ket{1}$ states of the first spin in the chain, and
that the same result can be obtained by encoding the information in
the \textit{sign} of the polarization using the states $\delta\rho_\pm
=\pm\sigma_z^1$.  This encoding is not enough, however, to transfer
quantum information: this would require the additional transfer of
information about the phase coherence of a state, for example by
transporting a state $\delta\rho_\pm =\pm\sigma_x^1$.  The problem is
that evolution of this state creates a highly correlated state, as
$\sigma_x^1$ evolves to $\prod_{i=1}^{n-1}\sigma_z^i\sigma_\alpha$,
where $\alpha=x(y)$ for $n$ odd (even). Information can be extracted
from this entangled state only with a measurement~\cite{DiFranco08c},
at the cost of destroying the initial state and of introducing
classical communication and conditional operations.

\begin{figure*}[t]
	\centering
\includegraphics[scale=0.65]{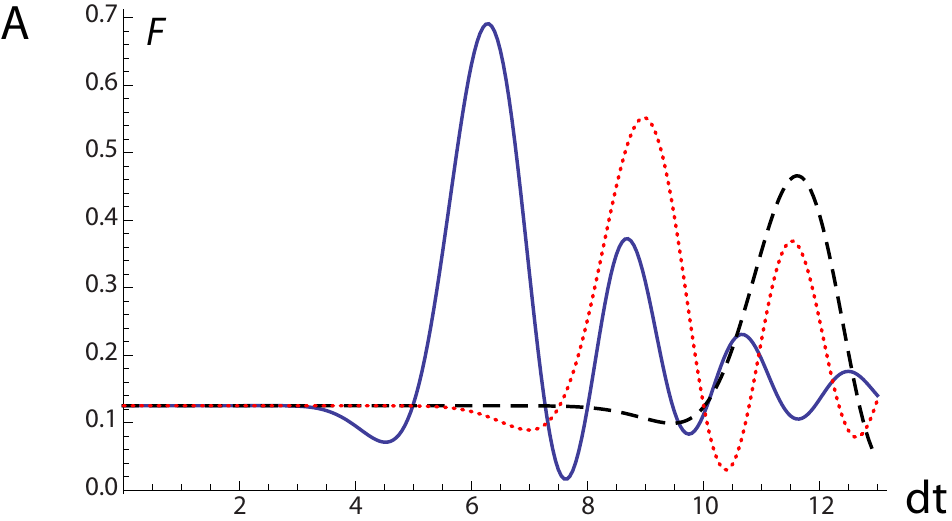}\ \ \ \ \ \ \ \  
\includegraphics[scale=0.65]{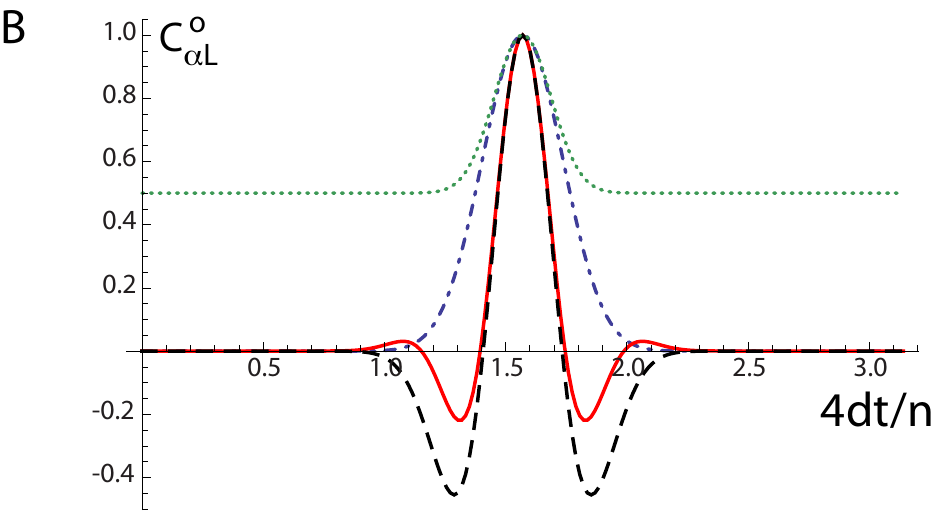}
\caption{(Color online) Transport of the four logical states as a
function of time (normalized by the coupling strength).  (A)
Entanglement fidelity, $F=\frac14\sum_{\alpha} C^h_{\alpha L}$, for
the transport under the homogeneous xx-Hamiltonian, for chains of
$n=10$ (blue, solid line), 15 (red, dotted line), and 20 (black,
dashed line) spins. (B) Transport of the four logical basis states
under the engineered optimal-coupling xx-Hamiltonian in a 20-spin
chain. $\sigma_{xL}$: Blue, dash-dotted line. $\sigma_{yL}$: Black,
dashed line. $\sigma_{zL}$: Red, solid line. $\sigma_{\Idf L}$: Green,
dotted line. }%
\label{fig:Engineer}%
\end{figure*}

A simple two-qubit encoding allows for the transport of a bit of
quantum information~\cite{Markiewicz09}. For evolution under the
xx-Hamiltonian, such encoding corresponds to the zero-eigenvalue
subspace of the operator $\sigma_z^1+\sigma_z^2$. A possible choice of
logical qubit observables is given by
\begin{equation}
	\begin{array}{ll}
	\sigma_{xL}^{\textsf{xx}}=(\sigma_x^1\sigma_x^2+\sigma_y^1\sigma_y^2)/2\
	\ &\ \
	\sigma_{yL}^{\textsf{xx}}=(\sigma_y^1\sigma_x^2-\sigma_x^1\sigma_y^2)/2\\
	\sigma_{zL}^{\textsf{xx}}=(\sigma_z^1-\sigma_z^2)/2\ \ &\ \
	\openone_L^{\textsf{xx}}=(\openone-\sigma_z^1\sigma_z^2)/2 ,
\end{array}
\label{eq:LogFF}
\end{equation}
which corresponds to an encoded pure-state basis $\ket{0}_L=\ket{01}$
and $\ket{1}_L=\ket{10}$.  If we perform the transport via the
dq-Hamiltonian, the required encoding is instead given by the basis
$\ket{0}_L=\ket{00}$ and $\ket{1}_L=\ket{11}$, as following from the
similarity transformation between xx- and dq-
Hamiltonians. Accordingly, the operator basis for the transport in
mixed states under the dq-Hamiltonian is
\begin{equation}
	\begin{array}{ll}
	\sigma_{L}^{\textsf{dq}}=(\sigma_x^1\sigma_x^2+\sigma_y^1\sigma_y^2)/2\
	\ &\ \
	\sigma_{yL}^{\textsf{dq}}=(\sigma_y^1\sigma_x^2-\sigma_x^1\sigma_y^2)/2\\
	\sigma_{zL}^{\textsf{dq}}=(\sigma_z^1-\sigma_z^2)/2\ \ &\ \
	\openone_L^{\textsf{dq}}=(\openone-\sigma_z^1\sigma_z^2)/2 .
\end{array}
\label{eq:LogDQ}
\end{equation}

We can calculate the transport functions $C_{\alpha L}(t)$,
$\alpha=\{x,y,z,\Id\}$, from the overlap of the evolved state with the
desired final state, for example for xx transport this yields
expressions of the form
\begin{equation}
	C_{yL}(t)=\tr{U_{\textsf{xx}}(t)\sigma_{yL}^{\textsf{xx}}U^\dag_{\textsf{xx}}(t)(\sigma_y^n\sigma_x^{n-1}-\sigma_x^n\sigma_y^{n-1})/2}.
\end{equation} 
For the homogenous xx-Hamiltonian we find 
\begin{equation}
C^h_{\Idf L}(t)=\half\left\{1+\left[A^{h}_{1,n-1}(t)A^{h}_{2,n-1}(t)-A^{h}_{1,n}(t)A^{h}_{2,n}(t)\right]^2\right\},
\end{equation}
\begin{equation}\begin{array}{ll}
C^h_{(x,y)L}(t)=&\frac{2(\pm1)^{n+1}}{(n+1)^2}\sum_{k,h} (-1)^{h+k} e^{i t \left(\omega _h\mp\omega _k\right)}\\ 
&\times\left[\sin (2 \eta ) \sin (\kappa )+\sin (\eta ) \sin (2 \kappa )\right]^2,
\end{array}
\label{eq:LogY}
 \end{equation}
\begin{equation}
C^h_{zL}(t)=\half\left[P^{h,\textsf{xx}}_{1,n}(t)-2P^{h,\textsf{xx}}_{1,n-1}(t)+P^{h,\textsf{xx}}_{2,n-1}(t)\right],
\end{equation}
where we have defined $\eta={\pi h}/({n+1})$.  Note that the same
expressions hold for the evolution of the states in
Eq.~(\ref{eq:LogDQ}) under the dq-Hamiltonian.  The transport under
this Hamiltonian is, however, imperfect, not only because the
transfer fidelity of each basis state is less than 1, but also
because the maximum values occur at slightly different times. In
Fig.~\ref{fig:Engineer} we plot the reduced entanglement
fidelity~\cite{Fortunato02b,NielsenFid} of such a transport process,
computed as $F(t)=\frac14\sum_{\alpha} C^h_{\alpha L}(t)$, for chains
of different lengths.

The transport of the logical states under the engineered Hamiltonian
${\cal H}^o_\textsf{xx}$ with optimal couplings is given by:
 \begin{equation}
 C^o_{\Idf L}(t)=\half \left[1+\sin( \tau) ^{4 (n-2)}\right],
\end{equation} 
 \begin{equation}C^o_{xL}(t)=\sin({\tau})^{2(n-2)},\end{equation} 
 \begin{equation}C^o_{yL}(t)=\sin(\tau)^{2 (n-2)} \left[1-2 (n-1) \cos ^2(\tau)\right], \end{equation}
\begin{equation}
\begin{array}{l}
C^o_z=\half \left\{\sin(\tau) ^{2 (n-3)} \left[(n-1)
\cos^2(\tau)-1\right]^2\right.\\ \left.+\sin(\tau)^{2 (n-1)}-2
(n-1)\cos^2(\tau) \sin(\tau)^{2 (n-2)}\right\} .
\end{array}
\end{equation} 

At the time $t^\star$ defined in Section \ref{Perfect} the basis
states are transported with fidelity one. It is then possible to
transfer an arbitrary state with unit fidelity
(Fig.  \ref{fig:Engineer}).  Note that because of the interplay of the
mirror inversion operated by the xx-Hamiltonian and the similarity
transformation between the xx- and dq-Hamiltonians, an additional
operation is needed to obtain perfect transport with the latter
Hamiltonian. Specifically, for chains with an even number of spins, a
$\pi$ rotation around the $x$-axis is required, which can be
implemented on the whole chain or on the last two spins encoding the
information. As this is a collective rotation, independent of the
state transported, arbitrary state transfer is still possible.

It is also worth noting that the above encoding protocol can be
extended to more than a single logical qubit, for example by encoding
an entangled state of two logical qubits into four
spins~\cite{Hodges07,Henry07}, such as an encoded Bell state
$\kpsi=(\ket{01}_L+\ket{10}_L)/\sqrt{2}$.  Provided that the extra
encoding overhead can be accommodated, this will in principle allow
perfect transport of entanglement through a completely mixed chain.

Altogether, these results point to a strategy for perfect transport in
spin wires, without the need of initialization or control, but only
exploiting control in a two-qubit (possibly four-qubit) register at
each end of the wire. The simplicity of such a protocol opens the
possibility for experimental implementations, as we proceed to discuss
next.

%====================================================
\section{Experimental Platforms}
%====================================================

\begin{figure*}[bht]
\centering
		\includegraphics[scale=0.65]{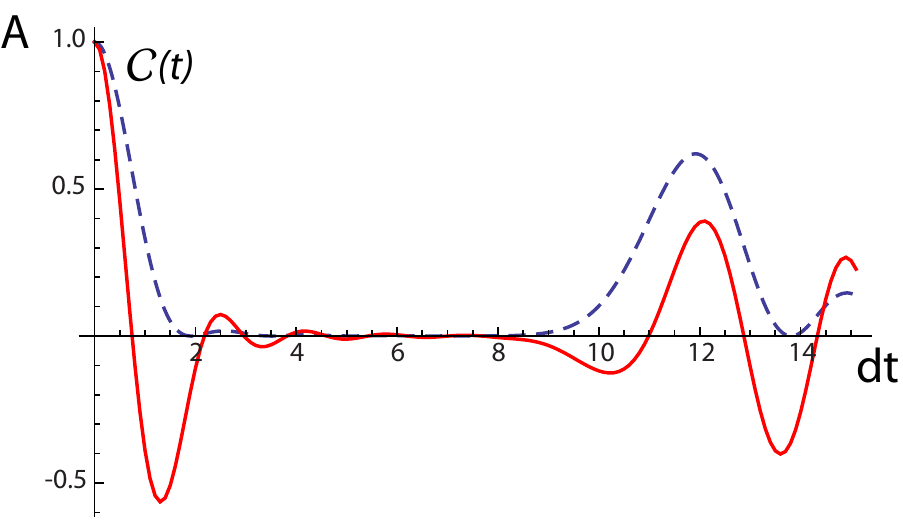}\ \ \ \ \ \ \ \  
		\includegraphics[scale=0.65]{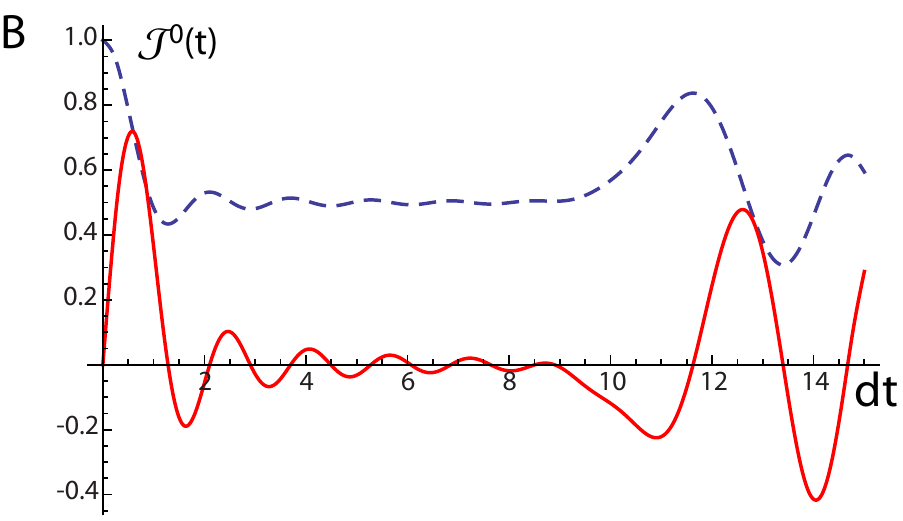}
\caption{(Color online) Transport of single-spin polarization
	$\delta\rho_z^{1,n}$ (blue, dashed) and logical-$y$ state
	$\delta\rho_y^L$ (red), Eq. \ref{eq:Expylog}, in a $n=21$-spin
	chain. (A) correlation of the evolved state with the initial
	state, $\mathcal{C}(t)=\tr{\rho(t)\rho(0)}$, which also
	indicates transport from one end of the chain to the other.
	(B) Zero quantum coherence intensities for the two initial
	states. }
\label{ShowMirror}
\end{figure*}

While many theoretical advances have been made in the study of
information transfer in spin chains, experimental implementations 
are still limited. Since departures from the idealized theoretical
models, due for instance to long-range couplings, the presence of a
bath, or variations in the coupling strengths, make real systems much
more complex to analyze analytically, experimental investigations able
to study these issues are needed. Studying quantum transport
properties in highly mixed spin chains thus serves a dual
purpose. First, the similarities of transport properties of pure and
mixed states makes the latter a good test-bed for experiment. Second,
protocols for perfect state transfer via mixed-state quantum wires
allow us to relax some of the requirements for QIP architectures.

Mixed-state spin chains are encountered in a number of physical
applications. Examples range from phosphorus defects in silicon
nanowires~\cite{Ruess07}, to quantum dots~\cite{Wang04,
Nikolopoulos04}, from polymers such as
polyacetylene~\cite{Nechtschein80} and other molecular
semiconductors~\cite{Petit90}, to solid state defects in diamond or
silicon carbide~\cite{Weber10,Wrachtrup10}.  In particular, the
completely mixed-state chain studied here, corresponding to the
infinite temperature limit, may often be a better approximation to the
thermal states of these systems than low-temperature thermal states
that may be viewed as perturbations to the ground state~\endnote{As it emerged from studies of coherence time with polarization~\cite{Fischer09,Takahashi08}, many properties do not depend linearly on the deviation from the pure state.}.

In what follows, we describe two experimental platforms that best
exemplify the advantages of transport via mixed-state chains.

%====================================================
\subsection{Simulations in solid-state NMR systems}
%=====================================================

Recently, nuclear spin systems in apatite crystals have emerged as a
test-bed to probe quasi-onedimensional (1D) dynamics, including
transport and
decoherence~\cite{Cho06,Cappellaro07a,Zhang09,RufeilFiori09,Ramanathan10}. Because
the nuclear spins in apatites are found in a highly mixed state at
room temperature, they are particularly well-suited for the
experimental study of the protocol for quantum information transport
outlined in the previous section.  NMR techniques enable this
exploration even in the absence of single-spin addressing and readout.

The crystal structure of fluorapatite [Ca$_5$(PO$_4$)$_3$F] and
hydroxapatite [Ca$_5$(PO$_4$)$_3$(OH)] presents a favorable geometry
where $^{19}$F or $^1$H nuclear spins are aligned in linear chains
along the crystal c-axis with inter-spin spacings much shorter than
the distance to other parallel chains.  In a sufficiently strong
magnetic field, the nuclear spins interact via the \textit{secular}
dipolar Hamiltonian~\cite{Munowitz87b}, 
\begin{equation}
\ham_{\textsf{dip}}=\sum_{j< l}^n d_{j l}
\left[\sigma_z^j\sigma_z^{ l} - \frac{1}{2}
(\sigma_x^j\sigma_x^{ l}+\sigma_y^j\sigma_y^{ l})\right],
\label{eq:dip}
\end{equation}
where the couplings depend on the relative positions as $d_{j
l}=(\mu_0/16\pi) (\gamma^2 \hbar /r_{j l}^3) (1-3\cos^2\theta_{j l})$,
with $\mu_0$ being the standard magnetic constant, $\gamma$ the
gyromagnetic ratio, $r_{j l}$ the distance between nucleus $j$ and
$l$, and $\theta_{j l}$ the angle between $\vec r_{j l}$ and the
$z$-axis, respectively.  The apatite geometry gives a ratio of
intra-chain to inter-chain couplings of about $40$, allowing the
evolution to approximate well the expected 1D dynamics over
sufficiently short time scales~\cite{Zhang09}.

Known pulse sequences~\cite{Munowitz87,Baum85,Ramanathan03} are able
to synthesize the dq-Hamiltonian from the secular dipolar
Hamiltonian. Furthermore, by relying on the symmetry breaking due to
defects and on incoherent control, we showed in
Ref. \cite{Cappellaro07a, Zhang09} how to prepare the initial state of
relevance for polarization transport,
$\delta\rho_z^{1,n}\propto\sigma_z^1+\sigma_z^n$ (notice that because
of the symmetries in the chain and control Hamiltonians, it is not
possible to prepare the state $\delta\rho_z^{1}\propto\sigma_z^1$).

Similar control protocols can be used to prepare other states for the
transport of quantum information.  Specifically, we want to prepare
states such as $\sigma_{xL}^{\textsf{dq}}$ and
$\sigma_{yL}^{\textsf{dq},L}$. To do so, one can first prepare the
state $\delta\rho_z^{1,n}$ and then let the system evolve under the
dq-Hamiltonian for a short time~\endnote{A time $t\approx 1.5/d$ is
found to maximize the fidelity of the prepared state.}. A so-called
double-quantum filter then selects the desired state
$\delta\rho_y^L\propto\left.\sigma_{yL}^{\textsf{dq},L}\right|_{1,2}+\left.\sigma_{yL}^{\textsf{dq},L}\right|_{n-1,n}$,
that is,
\begin{equation}
\delta\rho_y^L\propto(\sigma_y^1\sigma_x^2+\sigma_x^1\sigma_y^2)/2+(\sigma_y^{n-1}\sigma_x^n-\sigma_x^{n-1}\sigma_y^n)/2.
\label{eq:Expylog}
\end{equation}
Similarly, a $\pi/4$ collective rotation around $z$, prior to the
double-quantum filter, is needed to select the $\delta\rho_x^L$
operator. The double-quantum filter is a form of temporal
averaging~\cite{Knill98pp}, consisting in phase shifts of the pulse
sequences in successive experiments.  When averaging the experimental
results, such phase shifts cancel out contributions to the signal
arising from states outside the double-quantum coherence
manifold. Similar techniques are well established in
NMR~\cite{Rance83}, and have been used to study transport in
fluorapatite~\cite{Cappellaro07a}.

A suitable metric of transport would then be given by the correlation
of the evolved state with the initial state,
$\mathcal{C}(t)=\tr{\rho(t)\rho(0)}$, since this contains the usual
transfer terms (correlation of the evolved state with the desired
final state at the chain end,
$\tr{\left.\sigma_{yL}^{\textsf{dq},L}(t)\right|_{1,2}\left.
\sigma_{yL}^{\textsf{dq},L}\right|_{n-1,n}}$).  Even if the techniques
just outlined are able to prepare the desired initial state,
single-spin detection is not possible in conventional NMR, preventing
quantum information transport to be directly measurable. Still, there
exist other signatures that reliably indicate when the transport from
one end of the chain to the other has occurred. These signatures can
be extracted experimentally from the measurement of collective
magnetization, via so-called multiple quantum NMR
techniques~\cite{Baum85,Munowitz87b}. These techniques are extremely
useful to probe multi-spin processes and gain insight into many-body
spin dynamics~\cite{Baum85,Ramanathan03,Cho05,Cho06}, as they reveal
the multiple quantum coherence (MQC) intensities of a spin state, thus
effectively allowing a partial state tomography.

The $n${th} order MQC signal (when the observable is the total
magnetization $Z$) is given by
\begin{equation}
\begin{array}{l}
	\mathcal{J}^{n}_{\rho}(t)=\mathcal{F}_\varphi\{\textrm{Tr}\left[\right.e^{-i\varphi_n
	Z}U(t) \rho(0)U(t)^\dag e^{i\varphi_nZ} \\
	\quad\qquad\qquad\qquad\left.\times U(t)ZU(t)^\dag\right]\},
\end{array}
\end{equation}
where $\mathcal{F}_\varphi\{\cdot\}$ is the Fourier transform with
respect to the phase $\varphi$ and $U(t)$ is evolution under 
the dq-Hamiltonian~\cite{Ramanathan03}.  For an arbitrary initial
state $\rho(0)$, this corresponds to
\begin{equation}
	\mathcal{J}^{n}_{\rho}(t)=
\tr{\mathcal{P}_{n}[\rho_j(t)]~\mathcal{P}_{-n}[U(t)ZU(t)^\dag]},
\label{Jmqc}
\end{equation}
where $\mathcal{P}_n[\cdot]$ denotes the projector onto the $+n$
coherence manifold.

Although in 3D systems high coherence orders can be created, the 1D,
nearest neighbor dq-Hamiltonian creates only two-spin excited states
(zero and double quantum coherences), and thus it does not populate
higher coherence order manifolds~\cite{Feldman97}.  Furthermore, it
was observed in Ref. \cite{Cappellaro07l} that upon preparation of the
state relevant for transport, $\delta\rho_z^{1,n}$, the zero- and
double quantum intensities $\mathcal{J}_z^{0,2}(t)$ produced a clear
signature of the transport.

In the nearest-neighbor approximation, with 
$d=-{\mu_0\gamma^2\hbar}/({8\pi r_{nn}^3})$ and $r_{nn}$ being the
nearest-neighbor distance, the MQC intensities can be calculated
analytically, in the form
\begin{equation}
\mathcal{J}_z^{0,2}(t)=\frac{\alpha_{0,2}}{n+1} \sum_{k=1}^n
\sin^2(\kappa) \cos^2(2\omega_kt+\phi_{0,2}),
\end{equation} 
where $(\phi_0$$=$$0, \alpha_0$$=$$2)$ for the zero quantum and
$(\phi_2$$=$$\frac\pi2, \alpha_2$$=$$1)$ for the double quantum
intensities, respectively.  Similarly, we can calculate the MQC
intensities for the initial states corresponding to
$\delta\rho_{x,y}^L$ and evolving under the dq-Hamiltonian. Using the
transformation to Bogoliubov operators~\cite{Cappellaro07a}, we
obtain:
\begin{equation}
\mathcal{J}^{0,2}_{yL}(t)=\frac{\alpha_{0,2}}{n+1}\displaystyle\sum_{k=1}^n\sin(\kappa)\sin(2\kappa)
\sin\left(4\omega_kt+2\phi_{0,2}\right)
\label{eq:MQClog}
\end{equation}
whereas the state $\delta\rho_x^L$ gives a zero signal.

In figure \ref{ShowMirror}, we compare the transport metric
$\mathcal{C}(t)$ with the MQC intensities $\mathcal{J}^0(t)$. A
signature of transport from one end to the other of the chain is
apparent in the coherence intensities. The observed local maxima in
the MQC intensities at the \textit{mirror time} $t^*\sim n/(2d)$
\cite{Zhang09} is due to constructive interferences when the
propagation has traveled the length of the chain and is reflected
back~\cite{Ramanathan10}. The MQC signature would be amenable to
experimental tests in solid-state NMR systems, by following the
distinctive features of the MQC intensities evolution.  Other, more
comprehensive forms of state tomography~\cite{vanBeek05} inspired by
MQC techniques, could eventually be used to gather more information
about the evolved state.

%=======================================================
\subsection{A quantum computing architecture in diamond}
%=======================================================

We now turn to a promising implementation of the protocol for perfect
quantum information transfer described in Sec. \ref{qubit}.
Distributed quantum computing
schemes~\cite{Campbell08,Jiang08b,Burgarth06} could play an important
role in recently proposed solid-state quantum computing architectures
based on defects in diamond. The nitrogen-vacancy (NV) center in
diamond has emerged as an ideal qubit
candidate~\cite{Wrachtrup06,Childress06,Cappellaro09}, thanks to its
long coherence times and the possibility of optical initialization and
readout even at room temperature. This defect can be created by
implanting Nitrogen defects in diamond and allowing vacancies to
recombine with them at high temperature. While Nitrogen implantation
can be done with high
precision~\cite{Weis08,Toyli10,Naydenov10,Spinicelli10}, the Nitrogen
to NV conversion is limited.  The remaining Nitrogen defects (P1
centers~\cite{Hanson06}) are electronic spin 1/2 that can be used as
quantum wires to connect the NV-center qubits. While NV centers can be
initialized to their ground state and controlled individually by a
combination of microwave and optical control~\cite{Maurer10}, the P1
centers will be found in a highly mixed state and will only be able to
be controlled collectively.

The ideas developed in the previous sections find an ideal
implementation in this engineered QIP system.  Control on the NV
centers at each end of the chain allows to create the logical states
[Eq. (\ref{eq:LogDQ})] comprising the first neighboring P1 center
(notice that control on just the end chain spin could allow full
controllability of the chain~\cite{Fitzsimons06}, although this might
not be efficient~\cite{Burgarth10,Caneva09}).

The Nitrogens could be implanted at separations
$r_{i,i+1}=r_\text{min}\frac{\sqrt[3]{n/2}}{ \sqrt[6]{j (n-j)}}$, with
$r_\text{min}$ being the minimum separation, in such a way that the
couplings follow the ideal distribution that yields optimal
transport. Although the implantation precision is low at present,
technological advances should be able to reach the regime where the
transfer protocol becomes robust against errors in the coupling
strength~\cite{DeChiara05}. The P1 centers will then interact via the
dipolar interaction, which can be truncated to its secular part,
Eq. (\ref{eq:dip}), at high enough magnetic fields (in practice less
than $\approx\,100$ Gauss for a minimum distance between Nitrogens
$r_\text{min}\sim15$~nm, corresponding to a coupling strength of
$\approx\,15$kHz).

Using multiple pulse sequences~\cite{Ramanathan03}, the dipolar
Hamiltonian is modulated into the dq-Hamiltonian that we have shown
allows for perfect state transfer. At the same time, the pulse sequence
refocuses the hyperfine interaction with the Nitrogen nuclear
spin~\endnote{The refocusing is effective only in the presence of a large enough external magnetic field,  when the hyperfine interaction reduces to its secular component. Otherwise, other decoupling techniques should be used.} as well as the coupling to the quasi-static \carb nuclear
spin-bath. 
Assuming a 5\% error in the Nitrogen positioning, chains of $n\sim 15$
spins with minimum separation of 15nm would allow for information
transport in about $t^{\star}=200\mu$s, with high fidelity~\cite{DeChiara05,Kay06}.  Local operations at the NV center, enhanced
by a register of nuclear spins~\cite{Cappellaro09}, would allow for
quantum error correction, while the separation between NV centers
achieved thanks to the P1 wires would enable individual addressing of
the NV qubits by sub-diffraction-limit optical
techniques~\cite{Rittweger09,Maurer10}.  Ultimately, this scheme could
then serve as the basis for a scalable, room temperature quantum
computer.

%====================================================
\section{Conclusions}
%====================================================

We have investigated the properties of quantum information transport
in mixed-state spin chains. Focusing, in particular, on the infinite
temperature limit, we have identified strong similarities between
pure- and mixed- state transport. These similarities enable the
simulation of pure-state transport properties using more readily
accessible high-temperature mixed states. Specifically, we could apply
results derived for pure-state transport to achieve perfect state
transfer with an engineered xx-Hamiltonian. Other recently proposed
schemes, involving for instance weaker couplings of the chain
ends~\cite{Gualdi09} or modulation of an external bias magnetic
field~\cite{Alvarez10}, should be further explored to determine under
which conditions they could be extended to mixed-state chains with
different coupling topologies. More generally, it would be interesting
to investigate the wave-dispersion properties~\cite{Osborne04} of
mixed-state chains versus pure-state chains, as mixed-state systems
are ubiquitous in experimental implementations.

In this paper, we have discussed in particular a potential
experimental platform provided by apatite crystals controlled by NMR
techniques. Experimental simulations of pure-state transport would
allow exploring the effects of disordered and long-range couplings,
interaction with an environment, and other non-idealities that are
bound to appear in practical implementations and that are not amenable
to direct analytical and/or numerical studies.

Furthermore, it becomes possible to use known results of pure-state
transport to devise protocols for perfect spin transfer even using
highly mixed states. Specifically, we have showed that combining a
simple encoding of the transmitted state into one or more spin pairs
with engineered couplings in the chain allows for the perfect transfer
of quantum information and potentially of entanglement. An additional
advantage of mixed-state is that they enable transport of relevant
states via a non-excitation conserving Hamiltonian, the
dq-Hamiltonian, which can be obtained by coherent averaging from the
naturally occurring magnetic dipolar interaction. These results have
been combined to obtain a proposal for scalable quantum computation architecture
using electronic spin defects in diamond, which may be experimentally
viable with existing or near-term capabilities.

%====================================================
\section*{Acknowledgements}
%====================================================

It is a pleasure to thank Wenxian Zhang for early discussions on
transport in spin chains. P.C. gratefully acknowledges partial support
from the National Science Foundation under grant number DMR-1005926.
L.V. gratefully acknowledges partial support
from the National Science Foundation under grant number PHY-0903727.

%====================================================

\bibliographystyle{apsrev4P}
\input{TransportMixed2.7.bbl}
\end{document}